
\input vanilla.sty
\nopagenumbers

\def\Kosterlitztwo{1}
\def\Brown{2}
\def\Gerling{3}
\def\PaperZ3{4}
\def\Borgsdeux{5}
\def\BorgsMiracle{6}
\def\Bindun{7}
\def\Bind2{8}
\def\Baxter{9}
\def\Comment{10}
\def\Kosterlitz{11}
\def\Paperq10{12}
\def\BorgsJanke{13}
\def\FS{14}
\def\Novotny{15}
\def\Gupta{16}
\def\Birosa{17}
\def\Bergneuhaus{18}
\def\BBN{19}
\def\Widom{20}

\def\Z{2.1}

\def\cite #1 {~#1\ }
\def\ref #1 {Eq.(~#1)\ }
\font\tenbf=cmbx10

\font\ninerm=cmr9

\font\eightrm=cmr8
\font\eightit=cmti8

\TagsOnRight
\hsize=5.0truein
\vsize=7.8truein
\parindent=15pt
\baselineskip=10pt
\def\qed{\hbox{${\vcenter{\vbox{
    \hrule height 0.4pt\hbox{\vrule width 0.4pt height 6pt
    \kern5pt\vrule width 0.4pt}\hrule height 0.4pt}}}$}}
\line{\eightrm International Journal of Modern Physics C \hfil}
\line{\eightrm $\copyright$\, World Scientific Publishing Company \hfil}
\vglue 5pc
\baselineskip=13pt
\centerline{\tenbf NUMERICAL STUDY OF FINITE SIZE SCALING}
\centerline{\tenbf FOR FIRST ORDER PHASE TRANSITIONS\footnote"${*}$"
{\eightrm\baselineskip=10pt Talk presented at the Workshop
on DYNAMICS OF FIRST ORDER TRANSITIONS, HLRZ, Forschungszentrum J\"ulich,
Germany, June 1-3, 1992 }}
\vglue 24pt
\centerline{\eightrm Alain BILLOIRE}
\baselineskip=12pt
\centerline{\eightit Service de Physique Th\'eorique de Saclay\footnote"${+}$"
{\eightrm\baselineskip=10pt Laboratoire de la Direction des Sciences de
la Mati\`ere du CEA \hskip 7.5cm \vskip .5cm Saclay Preprint SPhT/92-097}}
\baselineskip=10pt
\centerline{\eightit 91191 Gif-sur-Yvette Cedex, France}
\vglue 16pt
\centerline{\eightrm ABSTRACT}
{\rightskip=1.5pc
 \leftskip=1.5pc
 \eightrm\baselineskip=10pt\parindent=1pc
I present results of simulations of the q=10 and q=20 2-d Potts
models in the transition region. The asymptotic finite size behavior
sets in only for extremely
large lattices. We learn from this simulation that  finite size scaling
cannot be used to decide that a transition is first order.
\vglue 5pt
\noindent
{\eightit Keywords}\/: First Order Phase Transitions; Monte Carlo.
\vglue 12pt}
\baselineskip=13pt
\line{\tenbf 1. Introduction \hfil}
\vglue 5pt
This talk addresses the question of deciding whether a transition is
first order or not, using Monte Carlo simulations.

$\bullet$ Very strong transitions are easy to detect. The system behavior
is very close to the infinite volume behavior.
Ergodicity is broken. Thermodynamical quantities are discontinous
at the transition point, with metastable branches.
A starting configuration half ordered, half disordered will relax to
very different states on both sides of the transition.

With extreme statistics, one is able to sample the coexisting phases
(although this may be forbiddingly  costly). The
time evolution of any thermodynamical quantity then shows
flip-flops between the phases, and the corresponding probability
distribution is made of well separated peaks.\looseness=-1

$\bullet$ In less clear cases, one must simulate systems of increasing
volumes $L^d$ and try to convince oneself that the above described very
large volume behavior is
approached. Let me use  the language of energy driven transitions in
what follows, and introduce the energy probability distribution
$P_L(E)$. In the transition region, it has two
peaks of heights $P^o_L$ and $P^d_L$, separated by a minimum of height
$P^{min}_L$.
For a first order transition, at fixed $P^o_L/P^d_L$, one has
$$ {1\over L^d} \ln{P^{min}_L\over P^o_L} \to 0 \hskip 1cm L \to \infty$$
This has been proposed by Lee and Kosterlitz\cite{\Kosterlitztwo} as
an indicator of
first order phase\vadjust{\vskip 0pt\penalty-5000\vskip 0pt}
 transitions (see also\cite{\Brown} \cite{\Gerling}).

Another class of indicators are moments of the distribution $P_L(E)$, that
goes to zero in the large volume limit, for all temperatures,
but at a first order transition point. An example is the energy fluctuation
$CV/L^d=\beta^2  ( <E^2> - <E>^2 )$, another is Binder's famous
cumulent  $BL={1\over 3}(1-<E^4> / <E^2>^2)$.
One can also plainly look at a plot of $P_L(E)$ and decide ``by eyes''
whether it approaches two delta functions as $L$ grows.

$\bullet$ The most sophisticated (and trustworthy) method
is finite size scaling. One insist in seeing the  finite size behavior
as predicted by the theory in the vicinity of a first order phase transition.
In that case only can one be pretty sure that the trend observed for lattices
of increasing sizes does  continue up to the thermodynamical limit.
One  insist on seeing
$$ {1\over L^d} \ln{P^{min}(L)\over P^o(L)} \sim A / L $$
$$CV_{max}/L^d  \sim CV^{(1)} + CV^{(2)} / L^d $$
where $CV_{max}$ is the maximum of the specific heat,
$$BL_{min}/L^d  \sim BL^{(1)} + BL^{(2)} / L^d $$
where $BL_{min}$ is the minimum of $BL$

My interest in first order phase transitions started with the
1988-1989 controversy over the order of the
deconfinement phase transition of pure $SU(3)$ gauge theory,
and the question whether
some modified 3-d $Z(3)$ Potts model has a first order transition or not.
Our data \cite{\PaperZ3} were very convincing showing that $BL_{min}$ has a non
zero large volume limit, however we did not observe the predicted $1/L^3$
finite size behavior (our data behave nearly like $1/L^2$).
This led us to investigate the 2-d q=10 Potts model, as an example of
a model with a strong first order transition, searching for
 the predicted finite size behavior.

\vglue 12pt
\line{\tenbf 2. Exact Results \hfil}
\vglue 5pt
Those have been obtained \cite{\Borgsdeux} for models that
can be represented by a
contour expansion with small activities, like\cite{\BorgsMiracle}
the q states Potts
model for large q. In such a case,  the partition function for a
$L^d$ lattice with periodic boundary conditions
can be written as
$$ Z(\beta,L)  =   e^{- L^d \beta f_d(\beta)} + q e^{- L^d \beta f_o(\beta)}
 +   {\cal O}(e^{-b L})  e^{- \beta f(\beta) L^d }
\hskip .8cm ;\  b > 0 \ \hskip 1.0 cm (2.1) $$
where $f_o(\beta)$ and $f_d(\beta)$ are smooth $L$ independent
functions. The free energy is $f(\beta) =
min \{ f_o(\beta) , f_d(\beta) \}$.
The phenomenological two gaussian peak model of the energy
probability distribution $P_L(E)$ introduced by K. Binder and
D. Landau \cite{\Bindun,\Bind2}
follows through inverse Laplace transform. The above exact result fixes
the relative weights of the two peaks: At the infinite volume limit
transition point, $\beta = \beta_t$, the
ordered and disordered peak weights are exactly in the ratio q to one.

The two gaussian peak model is not however a good
representation of $P_L(E)$ for all $E$'s. It
fails to describe the region between
the two peaks, and does not account for the observed L dependence of the
position of the two maxima of $P_L(E)$ when $\beta = \beta_t$.
This would require the understanding of the correction term in \ref{\Z}.

To the order in $1/L^d$ we consider, all quantities are expressed in terms of
$\beta_t$ and of the energies and specific heats of the two coexisting
phases. The transition temperature, $E_o$, $E_d$ and
the difference $C_o-C_d$  are known
exactly for the 2-d Potts models \cite{\Baxter}.
It follows from\ref{\Z} that the specific heat
$$CV=\beta^2 L^d ( <E^2> - <E>^2 )$$
has a maximum at
$$
\beta(CV_{max}) = \beta_t - {\ln q \over{E_d-E_o}}{1\over{L^d}}
+{\beta^{(2)}_{CV} \over L^{2 d}} + {\cal O} (1/L^{3 d}).
$$
The height of this maximum increases linearly with $L^d$
$$
CV_{max} = L^d  {{\beta_t^2}\over 4} (E_o - E_d)^2
+ CV^{(2)} + {\cal O} (1/{L^{d}}).
$$
whereas for fixed $\beta \neq \beta_t$, $CV(\beta)$ goes to a
constant, as $L$ goes to infinity.
One finds that $BL$ reaches a minimum equal
to \cite{\Comment,\Borgsdeux,\Kosterlitz}
$$
BL_{min} =  -  {(E_o^2 - E_d^2)^2 \over {12 (E_o E_d)^2}}
+ {BL^{(2)} \over L^{d}} + {\cal O} (1/L^{2d})
$$
at the point
$$
\beta(BL_{min}) = \beta_t -{\ln \bigl(q(E_o/E_d)^2\bigr)
\over{E_d-E_o}}{1\over L^d}
+{\beta^{(2)}_{BL} \over L^{2 d}} + {\cal O} (1/L^{3 d}).
$$

Expressions of the coefficients $\beta^{(2)}_{BL}$,  $BL^{(2)}$,
$\beta^{(2)}_{CV}$ and $CV^{(2)}$ as functions
of the $E_i$'s and $C_i$'s can be found in \cite{\Kosterlitz}.
Although the use of $BL_{min}$ as an indicator of the order of
phase transitions has been much publicized, $CV_{max}/L^d$ is as good an
indicator indeed.
The value of $BL$ depends on the
choice made of the arbitrary constant one can add to the definition of the
energy, this leads to introduce the quantity \cite{\Paperq10}
$$
U4 =  {<(E-<E>)^4> \over{ <(E-<E>)^2>^2}}
$$
which is independent of such a constant.
$U4$ is strictly larger than one, but at a first order transition
point, in the infinite volume limit. For large but finite volumes,
$U4$ reaches a minimum
$$
U4_{min} =    1 + { 8 (C_o + C_d) \over{ L^d  \beta_t^2  (E_o-E_d)^2}}
+ {\cal O} (1/L^{2 d}).
$$
at the point
$$
\beta(U4_{min}) = \beta_t -{\ln \bigl(q\bigr)
\over{E_d-E_o}}{1\over L^d}
+ {(C_o-C_d) (\ln^2(q)-8)\over{ L^{2 d} 2 \beta_t^2 (E_o-E_d)^3}}
+ {\cal O} (1/L^{3 d}).
$$
$BL_{min}$ (or $CV_{max}/L^d$) and  $U4_{min}$ are dual since $BL_{min}$
(or $CV_{max}/L^d$) going to zero means second (or higher order),
whereas $U4_{min}$ going to one means first order.

The above formulae for the extrema $CV_{max},\ BL_{min},\
U4_{min}$ and the corresponding effective $\beta$'s
have higher power law corrections that may hide the asymptotic
behavior on lattices  that can be simulated. In contrast,
the expressions for bulk averages evaluated at the (infinite volume
limit) transition point
$\beta=\beta_t$ do not have power law corrections,
as a consequence of Eq.{\Z}. The average energy
is given by
$$
E(\beta_t) = {E_d + q E_o \over{1+q}} + {\cal O}( e^{-b L})
$$
and the  value of the specific heat is
$$
CV(\beta_t) = { C_d + C_o q   \over { 1 + q }}
+ {L^d q \over (1+q)^2} (E_o-E_d)^2 \beta_t^2
+ {\cal O}( e^{-b L})
$$
The energy at $\beta_t$ does not depend on the lattice size, up
to  exponentially small corrections. This  provides
an efficient estimator of the transition temperature
\cite{\BorgsMiracle,\BorgsJanke}
by the following ``two-lattice method''.
One simulates lattices of increasing sizes
$L_1 <  L_2 < L_3 <. . $, and consider
$\beta_{eff}(L_i,L_{i+1})$,  the solution of the fixed point equation
$$
E_{L_i}(\beta) = E_{L_{i+1}}(\beta)
$$
for $i=1,2, \dots$. The estimate $\beta_{eff}(L_i,L_{i+1})$
converges towards $\beta_t$ with exponential pace.
The spectral density  (a.k.a Ferrenberg-Swendsen\cite{FS}, or reweighting)
method is invaluable  for locating extrema and zeros with Monte Carlo data
as input. It
allows to reconstruct the value of a thermodynamical average for
any $\beta$ from one  run performed at a given
$\beta_{M.C.}$ in the vicinity of the transition. It is well known
that this method has problems close to a second order point. With
moderate statistics, it predicts extraneous extreme for e.g. the
specific heat\cite{\Novotny}. This never occurred to us with first order
points.

To summarize, the extrema of $CV_{max}/L^d$, $BL_{min}$, and
$U4_{min}$ behave in the large volume limit like
$X^1+X^2/L^d+{\cal O}(1/L^{2d})$,
$CV(\beta_t)$ behaves like
$X^1+X^2/L^d + {\cal O}(e^{-b L})$.
The four different constants $\{X^1\}$ are exactly known. One
single unknown  parameter, e.g. the ordered specific
heats  $C_o$, fixes the four $X^2$'s.

\vglue 12pt plus 1pt minus 1pt
\line{\tenbf 3. Simulation of the q=10 model \hfil}
\vglue 5pt

We have performed  a $\approx$ 800 CRAY X-MP hours
simulation \cite{\Paperq10} of the $q=10$ 2-d Potts model, in
order to determine
how large $L$ has to be in order to see the asymptotic regime
described in  \cite{\Borgsdeux}. This
is a model with a strong, obvious, first order transition, with
\cite{\Gupta} a correlation length
$\xi(\beta_t^+) \sim 6$.
The precision of our data is better by more than one order of
magnitude than in   \cite{\Bind2}.
We have compared our results for the extrema of $CV/L^d$,
 $BL$ and $U_4$, and for the value $CV(\beta_t)$
with the large volume predictions.

We  simulated lattices up to $L=50$ (where the autocorrelation times
are $\tau_S
\approx .9 \ 10^6$ and $\tau_{NS} \approx 2.5 \ 10^6$). For all
four quantities, we  see
deviations from the $X^1+X^2/L^d$ limiting behavior. There is nothing to
worry about that, it  only
means that our precision is good. Really disturbing
however, is that these corrections do not seem to behave simply as
function of $L$, and are definitely not under control.
The values for the four slopes $\{X^2\}$ one would infer from our
data give inconsistent estimates of $C_o$. Three possible explanations are
i) \ref{\Z} is only proven in the large q limit, it may not hold
down to  $q=10$.
ii) Much larger lattices may be needed in order to
extract the true asymptotic behavior, although in our data  $P_L(E)$ has a
textbook first order shape.
iii) A programming error is always possible.

Before doing the simulation, we hoped that
$CV(\beta_t)$ would be asymptotic earlier than $CV_{max}$, since
corrections are ${\cal O}(e^{-b L})$. The data do not substantiate this
hope,  $CV(\beta_t)$ has  larger error bars, but does not seem to reach
its asymptotic behavior earlier. Note that the estimate of
$C_o$ we get from
$CV(\beta_t)$ is much higher than  the others.\looseness=-1

\vglue 12pt
\line{\tenbf 4. Simulation of the q=20 model using the
Multicanonical Algorithm \hfil}
\vglue 5pt

The conventional Metropolis (and Swendsen-Wang \cite{\Birosa})
algorithm suffers from exponential slowing down.
This makes simulations on lattices much larger than used to day,
impossible even with vastly more powerful computers.
It has been proposed by B.~Berg and T. Neuhaus \cite{\Bergneuhaus} to
perform the simulation with an  Hamiltonian designed in such a way that
$P_L(E)$ is very smooth
between $E_o$ and $E_d$, and to reweight the events when computing
expectation values. The  new ``multicanonical''
algorithm has only polynomial slowing down.

We \cite{\BBN} used this algorithm in order to simulate the q=20 Potts model.
It has a stronger first order transition and a smaller correlation length
than the q=10 model. This means that the large volume regime sets in
for smaller lattices for $q=20$ than for $q=10$.
We ran mainly on IBM RS6000 workstations,
and simulated lattices as large as  $38^2$. In contrast with the q=10
case, the values of the four slopes $\{X^2\}$ one infer from our
data give consistent estimates of $C_o$. As an example, Fig1 gives our
results for $CV_{max}/L^d$ together with the theoretical estimate
using the value $C_o = 5.2 \pm .2$, and Fig2 gives our results for
$CV(\beta_t)$ together with the theoretical estimate using the
same value for $C_o $. Note that $CV(\beta_t)$, reaches its asymptotic
behavior much earlier than $CV_{max}$, as predicted by the theory.
In conclusion the asymptotic behavior predicted by K. Binder, and
later proven by C. Borgs and R. Koteck\'y only sets in for very large
lattices.
The lattice size must fulfil the conditions $L >> \xi$, $L^{d-1} >>
1/A^{od}$ where  $A^{od}$ is the order-disorder surface tension
(If Widom's relation \cite{\Widom} holds this condition is equivalent
to the first one),
and $L^d >> C_o / (E_o-E_d)$, $L^d >> C_d / (E_o-E_d)$,
where $>>$ means five to ten
times larger. It is unfortunate that for such large systems, the
transition is blatantly first order.

\vglue 12pt
\line{\tenbf Acknowledgements \hfil}
\vglue 5pt
The material of this talk originates from work done with my
collaborators Bernd Berg, Tanmoy Bhattacharya, Sourendu Gupta,
Andrers Irb\"ack, Robert Lacaze, Andr\'e Morel, Thomas Neuhaus and
Bengt Petersson.

\vglue  10.5cm
\includegraphics{Fig1.ps}
Figure 1: $CV_{max}/L^2$ as a function of $1/L^2$ for the 2-d q=20
Potts model.
\vfill\penalty -5000\vglue 8.5cm
\includegraphics{Fig2.ps}
Figure 2: $BL_{min}/L^2$ as a function of $1/L^2$ for the 2-d q=20
Potts model.

\vglue 12pt
\line{\tenbf  References \hfil}
\vglue 5pt
\medskip
\ninerm
\baselineskip=11pt
\frenchspacing

\item{\Kosterlitztwo} J. Lee and J.M. Kosterlitz, Phys. Rev. Lett.
65 (1990) 137.

\item{\Brown} F.R. Brown and A. Yegulalp, Phys. Lett. A155
(1991) 252.

\item{\Gerling}  R.W. Gerling and A. H\"uller, Universit\"at
Erlangen Preprint 1992.

\item{\PaperZ3} A. Billoire, R. Lacaze, and A. Morel,
Nuclear Physics B 340 (1990) 542.

\item{\Borgsdeux} C. Borgs and R. Koteck\'y, J. Stat. Phys. 61 (1990) 79.

\item{\BorgsMiracle} C. Borgs,  R. Koteck\'y and S. Miracle-Sole,
J. Stat. Phys. 62 (1991) 529.

\item{\Bindun} K. Binder and  D.P. Landau, Phys. Rev. B30 (1984) 1477.

\item{\Bind2} M.S. Challa, D.P. Landau and K. Binder,
Phys. Rev. B34 (1986) 1841.

\item{\Baxter}
R.J. Baxter, J. Phys. A15, (1982) 3329;
R.J. Baxter, { Exactly Solved Models in Statistical Mechanics},
(Academic Press 1982).

\item{\Comment} A. Billoire, S. Gupta, A. Irb\"ack, R. Lacaze,
A. Morel and B. Petersson,  Phys. Rev. B42 (1990) 6743.

\item{\Kosterlitz} J. Lee and J.M. Kosterlitz, Phys. Rev. B43 (1991) 3265.

\item{\Paperq10} A. Billoire, R. Lacaze, and A. Morel,
Nucl. Phys. B 370 (1992) 773.

\item{\BorgsJanke} C. Borgs and W. Janke, Phys. Rev. Lett. 68 (1992) 1738.

\item{\FS} A. M. Ferrenberg and R.H. Swendsen, Phys. Rev. Lett. 61 (1988) 2635;
63 (E) (1989) 1658.

\item{\Novotny} E.P. M\"unger and M.A. Novotny, Phys. Rev. 43B (1991) 5773.

\item{\Gupta} S. Gupta and A. Irb\"ack, Phys. Lett. B286 (1992) 112.

\item{\Birosa} A. Billoire, S. Gupta, A. Irb\"ack, R. Lacaze, A. Morel
and B. Petersson, Nucl. Phys. B358 (1991) 231.

\item{\Bergneuhaus} B. Berg and T. Neuhaus,  Phys. Lett. B 267 (1991) 249,
Phys. Rev. Lett. 68 (1992) 9.

\item{\BBN} B. Berg, A. Billoire and T. Neuhaus,  work in  slow  progress.

\item{\Widom} See e.g. D. B. Abraham in ``Phase Transitions and
Critical Phenomena'', Vol 10.\looseness=-1

\vfil\supereject
\bye